\documentclass[12pt]{article}
\usepackage{epsfig}
\usepackage{amsmath}
\usepackage{amsfonts}
\usepackage{amssymb}
\usepackage{cite,graphicx}

\newcommand{\pmi}{\text{\tiny L,R}}

\newcommand{\ir}{\text{\tiny IR}}
\newcommand{\uv}{\text{\tiny UV}}
\newcommand{\ri}{\text{\tiny R}}
\newcommand{\li}{\text{\tiny L}}
\newcommand{\ai}{\text{A}}
\newcommand{\sm}{\text{\tiny SM}}

\newcommand{\dark}{\text{\tiny dark}}
\newcommand{\dm}{\text{\tiny DM}}
\newcommand{\be}{\begin{equation}}
\newcommand{\ee}{\end{equation}}
\DeclareMathOperator{\Ei}{Ei}
\DeclareMathOperator{\sgn}{sgn}

\begin{document}
\pagestyle{empty}

\begin{center}
{\LARGE \bf  A Warped Model of Dark Matter}

\vspace{1.0cm}

{\sc Tony Gherghetta}\footnote{E-mail:  tgher@unimelb.edu.au}
{\footnotesize\sc and }
{\sc Benedict von Harling}\footnote{E-mail:  bvo@unimelb.edu.au}\\
\vspace{.5cm}
{\it\small {School of Physics, University of Melbourne, Victoria 3010,
Australia}}
\end{center}

\vspace{1cm}
\begin{abstract}
We present a model of dark matter in a warped extra dimension 
in which the dark sector mass scales are naturally generated without 
supersymmetry. The dark force, responsible for dark matter annihilating 
predominantly into leptons, is mediated by dark photons that naturally 
obtain a mass  in the GeV range via a dilaton coupling. As well as 
solving the gauge hierarchy problem, our model predicts dark matter in the 
TeV range, including naturally tiny mass splittings between pseudo-Dirac states. 
By the AdS/CFT correspondence both the dark photon and dark matter are 
interpreted as composite states of the strongly-coupled dual 4d theory. 
Thus, in our model the dark sector emerges at the TeV scale from the 
dynamics of a new strong force.

\end{abstract}

\vfill
\begin{flushleft}
\end{flushleft}
\eject
\pagestyle{empty}
\setcounter{page}{1}
\setcounter{footnote}{0}
\pagestyle{plain}

\section{Introduction} 
\label{secIntro}

An outstanding problem in particle physics and cosmology is to explain the origin 
of dark matter. So far our only hints of dark matter have been unable to shed much 
light on its nature and interactions. However, recently a number of anomalies from 
observational astrophysics have been reported \cite{atic,pamela,fermi,hess}.
Although these signals may find an astrophysical explanation \cite{Aharonian:1995zz,Hooper:2008kg},
an exciting possibility is that they are due to the annihilation of dark matter in the galactic halo. 
This would suggest a dark matter sector unlike any of the usual WIMP scenarios. 
According to Arkani-Hamed, Finkbeiner, Slatyer and Weiner \cite{dmref} (see  
\cite{exciting1a,Pospelov:2007mp,exciting2,Pospelov:2008jd} for related and earlier work), the astrophysical anomalies imply 
a dark matter sector containing a WIMP with mass near the TeV scale that 
predominantly decays into leptons via a dark force.

An attractive feature of this model is that it is able to reconcile and incorporate all of
the recent observations.\footnote{Various constraints on models of this type can be found e.g.~in \cite{kp,cdfgw,Galli:2009zc,
Bergstrom:2009fa,mpsv,Slatyer:2009yq, Cholis:2009gv, Dent:2009bv, Zavala:2009mi,Feng:2009hw, Buckley:2009in,Cirelli:2009dv,Papucci:2009gd}.}
This is done by introducing a hierarchy of scales to explain the different observations.
The excess in the combined flux of electrons and positrons, which was observed by FERMI and HESS 
\cite{fermi,hess}\footnote{These experiments do not confirm an edge feature in the spectrum 
claimed by ATIC~\cite{atic} but still find an excess.}, indicates a dark matter mass in the TeV range. 
The annihilation cross section required to obtain these signals and the signal observed by 
PAMELA~\cite{pamela} is much larger than that expected for a thermal WIMP. 
This requires a light dark photon which can increase the dark matter annihilation cross section
in the galactic halo via the Sommerfeld effect. Moreover, the dark matter must predominantly 
annihilate to leptons in order to explain the absence of an antiproton excess at PAMELA. 
This can be achieved when the annihilation is dominantly a two-step process:  The dark matter  first 
annihilates to dark photons and the latter subsequently decay to charged Standard Model particles 
via kinetic mixing with the Standard Model photon. If the mass of the dark photon is sufficiently small, 
only decays to leptons are kinematically allowed.

There are other experiments which may have found 
hints on the nature of dark matter: The direct detection experiment DAMA/LIBRA \cite{dama} observes an annual modulation 
which is consistent with dark matter scattering off nuclei. Inelastic scattering between two states with a mass 
split of order 100 keV can reconcile this signal with the null results of other direct detection 
experiments \cite{inelastic}.\footnote{Constraints on the inelastic scenario can be found e.g. in \cite{inelastic,MarchRussell:2008dy}.} Furthermore, the 511 keV line coming from the centre of the
galaxy, recently remeasured by the satellite INTEGRAL \cite{integral}, can be 
explained with a mass split of order 1 MeV and the decay of the heavier, excited state to the lighter 
state under the emission of electrons and positrons~\cite{exciting1a,exciting1b}.

It is a challenge to construct dark matter models which contain these features within this range of mass scales while simultaneously solving the hierarchy problem in the Standard Model. Nevertheless supersymmetric models have been proposed which can incorporate the essential features \cite{Ahw,Baumgart:2009tn,ks,crwy,mpz}. They all require a departure from the usual neutralino scenarios in the supersymmetric Standard Model. Other aspects of models along the lines of \cite{dmref} were discussed in \cite{Chen:2009dm,Chen:2009ab}.

Alternatively in this paper, within the context of solving the gauge hierarchy problem, we propose a 
nonsupersymmetric model of dark matter.\footnote{Sommerfeld enhancement mediated by the radion in a warped model has been studied in \cite{Agashe:2009ja}. In \cite{McDonald:2010iq}, light gauge bosons were obtained from an additional slice of AdS$_5$ with an IR scale of order GeV.} We will use the Randall-Sundrum framework~\cite{rs1} in order to 
generate the various mass scale hierarchies. The infrared (IR) scale is taken to be of order the TeV scale in order to solve the gauge hierarchy problem in the visible sector (the usual Standard Model). The dark sector is then modeled by introducing a dark photon and Dirac fermion into the warped extra dimension. A coupling to a dilaton background is 
used to localize the dark photon near the IR brane, which obtains a GeV scale mass when the dark gauge group,
U(1)$^\prime$, is broken on the ultraviolet (UV) brane. Thus, the GeV mass scale is naturally generated via the wavefunction overlap in the 5th dimension. 

The Dirac fermion is charged under the dark gauge group U(1)$^\prime$. A Majorana mass on the UV brane causes the massless fermion mode to decouple, leaving a pair of TeV-scale Majorana fermions with a mass splitting of order the MeV (or 100 keV) scale. The lightest fermion is stable, electromagnetically neutral and plays the role of the dark matter. 
The dark photon couples dominantly off-diagonal to the pair of pseudo-Dirac states, thereby naturally incorporating
the inelastic dark matter scenario~\cite{inelastic}. Furthermore, even though the effective five-dimensional (5d) coupling increases towards the UV brane, the Kaluza-Klein modes are always weakly coupled.

Our model of the dark sector is simple and economical, depending on effective bulk mass parameters which correspond to turning on operators of various dimensions in the corresponding four-dimensional (4d) dual theory. 
In fact, via the AdS/CFT correspondence~\cite{Maldacena:1997re,ahpr}, both the dark photon and dark matter are interpreted as composite states of the strongly-coupled dual theory. Moreover, unlike other proposals, the symmetry breaking sector is particularly simple, relying on the dark gauge group being broken by boundary conditions at the Planck scale. This means our dark sector is in fact Higgsless with there being no need to introduce a dark Higgs sector at low energies. Since the underlying dynamics is due to a strongly-coupled gauge theory, there is the possibility of heavier resonances being produced at the LHC.

\section{The Dark Gauge Boson} \label{sec5dmodel}

\subsection{A light gauge boson from localization}
\label{lightgeneral}
In order to generate the required mass scales in the dark matter sector we will consider a warped extra 
dimension~\cite{rs1}. The background geometry is assumed to be a slice of 5d AdS space with the metric
\begin{equation}
ds^{2}=g_{MN}dx^M dx^N=e^{-2ky} \eta_{\mu\nu}dx^{\mu}dx^{\nu}+dy^2~,
\end{equation}
where $k$ is the AdS curvature scale and $\eta_{\mu\nu}={\rm diag}(-1,+1,+1,+1)$ is the Minkowski metric. 
The 5th coordinate $y$ extends from $0$ to $L$. The boundaries at $y=0$ and $y=L$ are the locations of 
two 3-branes, the UV and IR brane, respectively.

The scale of the UV brane is $m_\uv =k$, where $k$ is taken to be of order the Planck scale, while the IR brane scale is chosen to solve the hierarchy problem. Just like in the original model~\cite{rs1} we assume that the Higgs field is confined to the IR brane with an IR scale $m_\ir \equiv k e^{-kL} \simeq$ TeV. We will also assume that the bulk contains gauge fields~\cite{dhr,pomarol} and fermions~\cite{gn,gp1} in order to explain the fermion mass hierarchy. Within this framework we will in addition introduce the dark sector fields.

\begin{figure}[t]
\begin{minipage}[b]{0.47\linewidth}
\centering
\includegraphics[scale=1.17]{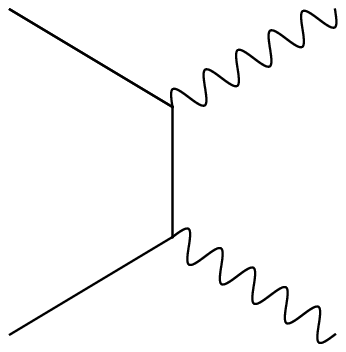}
\caption{Annihilation of dark matter to dark gauge bosons}
\label{annihilationFIG}
\end{minipage}
\hspace{0.6cm} 
\begin{minipage}[b]{0.477\linewidth}
\centering
\includegraphics[scale=1.2]{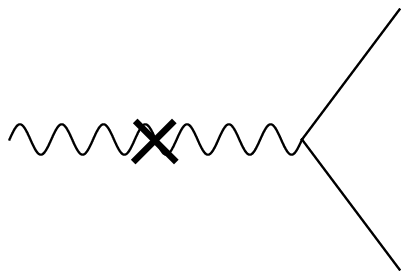}
\vspace{.4cm}
\caption{Decay of a dark gauge boson via kinetic mixing with the SM photon}
\label{decayFIG}
\end{minipage}
\end{figure}

In secluded models \cite{exciting1a,Pospelov:2007mp,exciting2,dmref,Pospelov:2008jd}, the dark matter does not couple directly to the 
Standard Model (SM). Instead, it is charged under a dark gauge group and the dark gauge bosons couple to the electromagnetic current via kinetic mixing with the photon.\footnote{This mixing stems from a mixing with the hypercharge gauge boson. Another possible mediator is a scalar which couples to the dark matter and which mixes with the SM Higgs (see also \cite{Bai:2009ms}).} If the dark gauge group is broken, the annihilation of dark matter to the SM is dominantly a two-step process: The dark matter annihilates to the dark gauge bosons via the process shown in Fig.~\ref{annihilationFIG}. The latter subsequently decay to light SM fermions (see Fig.~\ref{decayFIG}) and thereby can give rise to the signals observed by PAMELA, FERMI and HESS. In order to kinematically forbid decays to antiprotons, the dark gauge bosons must be lighter than 1 GeV. If they are too light, however, the Sommerfeld-enhanced annihilation of the dark matter in the early universe leads to an overproduction of gamma rays, in conflict with experimental bounds \cite{kp}. In Ref.~\cite{dmref}, the mass of the dark gauge bosons is therefore chosen to be $\cal O$(GeV).

We will now show how to obtain such an $\mathcal{O}$(GeV) gauge boson in a warped extra dimension when the
scale of the IR brane is associated with the TeV scale. The dark matter is discussed in Sect.~\ref{secdark}. We consider a bulk U(1)$^\prime$ gauge group and assume that the 
U(1)$^\prime$ is broken on the UV brane. In the following, we refer to the lightest Kaluza-Klein (KK) mode of the corresponding gauge field as the dark photon, $\gamma'$. If this mode is localized towards the IR brane, it obtains only a small mass from the symmetry breaking due to its small overlap with the UV brane.\footnote{Note that the U(1)$^\prime$ gauge group could also have been broken on the IR brane which would yield a small mass for a dark photon with a UV localized profile. However, as we will see later, in this case it is no longer possible to induce small mass splittings for the dark matter fermion by adding a boundary Majorana mass term.} In order to localize the dark photon, we follow \cite{Kehagias:2000au} and consider a (dimensionless) scalar field $\phi$ and the action
\be
\label{dilatoncoupling}
S^{(A')} \, = \, \int d^5x \sqrt{-g} \left[ -\frac{1}{4} e^{-2\phi} F'_{MN} F^{\prime MN}\right]\,.
\ee
We refer to $\phi$ as the dilaton due to its dilaton-like coupling to the gauge field strength, although 
we emphasize that we do not think of $\phi$ as the dilaton of string theory. In particular, we assume that the factor 
$\smash{e^{-2\phi}}$ only enters the Lagrangian of the dark matter sector. We assume that $\phi$ has 
a vacuum expectation value (vev) $\langle \phi \rangle$ which is varying along the 5th dimension. If the U(1)$^\prime$ is unbroken, corresponding to Neumann boundary conditions at the two branes, the equation of motion allows for a massless 4d mode with a constant profile, $f_\ai^{(0)}(y) = N_\ai^{(0)}$. 
The action for this mode then reads
\be
\label{ZeroModeAction}
(N_\ai^{(0)})^2 \, \int dy \, e^{-2 \langle \phi \rangle}  \int d^4x  \left[ -\frac{1}{4} F^{\prime(0)}_{\mu\nu} 
F^{\prime(0)\mu\nu} \right] \, .
\ee
From this, we see that the profile of the massless mode with respect to a flat metric is ${\widehat f}_\ai^{(0)}(y) \propto e^{-\langle \phi \rangle}$. For different vevs $\langle \phi \rangle$, the massless mode can be localized anywhere in the bulk. In particular, if the U(1)$^\prime$ is broken on the UV brane, the mass of the lightest KK mode can be made arbitrarily small by localizing the mode towards the IR brane.

\subsection{Linear dilaton profile}
\label{linDilaton}
We will now demonstrate the existence of such a light state in the spectrum quantitatively. For simplicity, we first consider vevs which are linear in the coordinate $y$, $\langle \phi \rangle = a -b k y$, where $a,b$ are constants and the relative sign is chosen for later convenience.  For a linear dilaton vev, the spectrum of KK modes of the dark gauge boson is straightforward to determine. However, even though we assume that the backreaction of the vev on the metric can be neglected, we will not try to find an action which dynamically generates such vevs. Instead in Sect.~\ref{expDilaton}, we will discuss dilaton vevs which are exponential in $y$ and show how to generate such profiles dynamically.

As noted in \cite{Tachibana:2001up} (see also \cite{Batell:2006dp}), by absorbing the linear dilaton vev into the 
gauge field, $\widehat{A}_M \equiv e^{-\langle\phi\rangle} A'_M$, the action becomes that of a `standard' gauge field in AdS$_5$, but with specific bulk and boundary masses.\footnote{Instead of an exponential prefactor $e^{-2\phi}$ (as in \eqref{dilatoncoupling}) with a linear vev, we could consider a linear coupling $\phi  F^{\prime 2}_{MN}$ with an exponential vev. Absorbing the vev into the gauge field, $\smash{\widehat{A}_M \equiv \langle \phi \rangle^{1/2} A'_M}$, again leads to a `standard' action for  $\smash{\widehat{A}_M}$. Fluctuations $\delta \phi$ of the dilaton couple in this frame via the term $\smash{\langle \phi \rangle^{-1}\widehat{F}_{MN}^2 \delta \phi}$ (plus additional terms involving derivatives of $\langle \phi \rangle$). As we will discuss in Sect.~\ref{CouplingSection}, in the cases of interest to us, the prefactor of the gauge kinetic term becomes exponentially small close to the UV brane. Accordingly, the coupling $\smash{\langle \phi \rangle^{-1}}$ of the dilaton to the gauge field becomes exponentially large in that region. We would therefore have to check whether $\delta \phi$ is strongly coupled to $\widehat{A}_M$ (cf.~Sect.~\ref{CouplingSection}). With an exponential prefactor as in \eqref{dilatoncoupling}, this issue can be avoided.}
 This simplifies the KK analysis. We work in the unitary gauge where the massive KK modes of $A'_5$ become the longitudinal modes of the massive 4d gauge bosons. Furthermore, we always choose boundary conditions on $A'_5$ such that this field has no massless mode. We will thus set $A^\prime_5 = 0$. Ignoring fluctuations $\delta \phi$ of the dilaton, the action then reads
\begin{multline}
    S^{(A')} = \int d^5x \, \left[-\frac{1}{4}\widehat{F}_{\mu \nu}^2-\frac{1}{2} e^{-2ky} \,   \left(\partial_y \widehat{A}_\mu\right)^2-\frac{1}{2} e^{-2ky} \, (b^2-2b) \, k^2  
    \widehat{A}^{2}_\mu \right.\\  \left.- \,e^{-2ky} \, b k  	\widehat{A}^{2}_\mu \,\bigl(  \delta(y) -\delta(y-L) \bigr)\right] \, .
\label{dphaction}
\end{multline}
This action was previously considered in \cite{Ghoroku:2001zu,Kogan:2001wp,bg} as a means to localize gauge fields. We obtain the boundary conditions on an orbifold by choosing $A^\prime_\mu$ (and thus $\smash{\widehat{A}_\mu}$) to be even under $\mathbb{Z}_2$. This leads to modified Neumann conditions, $\smash{(\partial_y -b k) \widehat{A}_\mu |_{0,L} =0}$, at the two branes. Alternatively, if the theory lives on an interval, the variation of the action at the two branes must vanish. The above boundary conditions are seen to be an admissible choice \cite{Hebecker:2001jb,Csaki:2003dt,Csaki:2003sh} in this case. They allow for a massless KK mode with profile $\smash{{\widehat f}^{(0)}_\ai \propto e^{b k y}}$. This profile agrees with that derived from \eqref{ZeroModeAction} and is UV (IR) localized for $b <0\, (b >0)$.

To break the gauge symmetry, we now impose the Dirichlet condition $\smash{\widehat{A}_\mu |_{0} = 0}$ (instead of the modified Neumann condition) at the UV brane. 
Thus, the dark sector is Higgsless in analogous fashion to Higgsless models of electroweak-symmetry breaking~\cite{Csaki:2003dt,Csaki:2003sh,higgsless}. Decomposing the gauge field as 
\be
\smash{\widehat{A}_\mu (x^\nu,y) = \sum_{n=0}^{\infty} \widehat{A}^{(n)}_\mu(x^\nu) \, {\widehat f}_\ai^{(n)}(y)}~,
\ee
and imposing the boundary conditions, leads to the solution 
\begin{equation}
 {\widehat f}_\ai^{(n)}(y)=\widehat{N}_\ai^{(n)} e^{ky}\left[
J_{b-1}\left(\frac{m_n}{k}e^{ky}\right)-\frac{J_{b-1}(\frac{m_n}{k})}{Y_{b-1}(\frac{m_n}{k})}
\; Y_{b-1}\left(\frac{m_n}{k}e^{ky}\right)\right]~,
\label{massive}
\end{equation}
where $\smash{\widehat{N}_\ai^{(n)}}$ is a normalization constant. The KK mass spectrum is determined from the algebraic equation
\begin{equation}
\frac{J_{b}\bigl(\frac{m_n}{m_\ir}\bigr)}{Y_{b}\bigl(\frac{m_n}{m_\ir}\bigr)} \, = \, \frac{J_{b-1}\bigl(\frac{m_n}{k}\bigr)}{Y_{b-1}\bigl(\frac{m_n}{k}\bigr)} ~,
\label{massquantization0}
\end{equation}
and expanding \eqref{massquantization0} for $m_n \ll m_\ir \ll k$, we find that the lowest mode obtains the mass
\begin{equation}
    m_0 \approx\,  e^{(1-b) \, k L}\; m_\ir~.
    \label{masseqn}
\end{equation}
If $b>1$, this mass will be sufficiently suppressed below the IR scale. Assuming $k = 10^{18}$ GeV and $m_\ir=1$ TeV (and thus $kL \simeq 34.54$), a dark photon mass of $\cal O$(GeV) is obtained for $b\simeq 1.2$. 

The wavefunction of the dark photon can be simplified by expanding \eqref{massive} for $\smash{\frac{m_0}{k}e^{ky}} \ll 1$ (which is satisfied everywhere between the UV and IR brane) and is, to a good approximation, given by the wavefunction of the former massless mode. Normalizing the wavefunction to obtain a canonical kinetic term, we find 
\be
\label{lightwavefunction}
 {\widehat f}_\ai^{(0)}(y) \, \simeq \, \sqrt{2 b k} \, e^{bk(y-L)}~.
\ee

Notice that the constant part $a$ of the dilaton vev has not played a role yet. The dilaton factor in \eqref{dilatoncoupling} is related to the inverse gauge coupling. As we want the dark photon to couple with $\mathcal{O}$(1) strength to charged matter in the IR, this factor should be $\mathcal{O}$(1) in the IR as well (see Sect.~\ref{CouplingSection}).
In the following, we will therefore focus on dilaton vevs which are small in the IR, leading to an $\mathcal{O}$(1) prefactor of the gauge kinetic term in that region. For simplicity, we will set $\smash{\langle\phi\rangle |_{y=L}=0}$, corresponding to 
$a = bkL$.

\subsection{Exponential dilaton profile}
\label{expDilaton}
We now consider the following action for the dilaton:
\be
\label{dilatonaction}
S^{(\Phi)} \, = \, - \frac{1}{2}\int d^5x \sqrt{-g} \left[ \partial_M \Phi \partial^M \Phi + m_\Phi^2 \Phi^2 + \delta(y) \, \lambda_0  \left(\Phi^2 - v_0^2 \right)^2 \right]\, ,
\ee
where we have included a UV boundary potential for the dilaton. The parameters $\lambda_0$ and $v_0$ have mass dimensions $-2$ and $\smash{\frac{3}{2}}$, respectively. The field $\Phi$, which has the canonical mass dimension, is related to the dimensionless field $\phi$ by $\Phi = f_\Phi^{3/2} \phi$, where the mass scale $f_\Phi$ is the 
dilaton decay constant. We assume that this scale is smaller than the 5d Planck scale $M_5$, 
where for example, it could be identified with the AdS scale $k$. 

The potential localized on the UV brane is consistent with the boundary condition $\smash{\partial_y\Phi - \lambda_0 \Phi (\Phi^2-v_0^2)|_0 = 0}$. This causes the dilaton to obtain a vev. As will be discussed below, we restrict ourselves to regimes in which the backreaction of this vev on the metric can be neglected. The bulk equations of motion are solved by \cite{Goldberger:1999uk}
\be
\langle \Phi (x^\nu,y) \rangle \, = \, C_1 e^{(4+\alpha)ky} \, + \, C_2 e^{-\alpha ky} \, ,
\ee
where, for later convenience, we have defined $\smash{\alpha\equiv \sqrt{4 + m_\Phi^2/k^2}-2}$.
At the IR brane, we impose the Neumann condition $\smash{\partial_y\Phi|_{L} = 0}$.
This fixes the vev up to an overall constant $v$:
\be
\label{dilatonvev}
\langle \Phi (x^\nu,y)\rangle \, = \, v \left( e^{-\alpha ky} \, + \, \frac{\alpha}{4+\alpha} \, e^{-2(2+\alpha) kL} \, e^{(4+\alpha)ky} \right) \, \simeq \, v \, e^{-\alpha ky} \, .
\ee
We restrict ourselves to the case $\alpha>0$ (corresponding to $m_\Phi^2>0$). In the last step in \eqref{dilatonvev}, we have ignored the term 
$\smash{\propto e^{(4+\alpha)ky}}$ which becomes relevant only close to the IR brane.\footnote{We see that the vev is peaked at one brane. The contribution of the vev to the energy density after compactifying to 4d has therefore a negligible dependence on the size of the extra dimension. Accordingly, the dilaton does not interfere with
the stabilization of the radion, for example via the Goldberger-Wise mechanism \cite{Goldberger:1999uk}.} Using this approximation in the boundary condition at the UV brane, we find
\be
v\, \simeq \,\pm \left( v_0^2 \,- \, \frac{\alpha k}{ \lambda_0}\right)^{1/2} \,.
\ee 
In the following, we assume that $v >0$ and $\smash{\alpha  k/ \lambda_0 \ll v_0^2}$ so that $v\simeq v_0$. From \eqref{dilatonvev}, we see that we can neglect the backreaction 
of the vev on the metric as long as $\smash{ v \ll  M_5^{3/2}}$. The prefactor of the dark gauge field strength in \eqref{dilatoncoupling} becomes
\be
\label{prefactor}
e^{-2\langle\phi\rangle}=e^{- \beta \, e^{-\alpha ky}} \, ,
\ee
where we have defined $\smash{\beta \equiv 2 v/f_\Phi^{3/2}}$. We will be interested in $\alpha={\cal O}(1)$ and $\beta={\cal O}(10)$. The prefactor in this case quickly goes to 1 away from the UV brane. In particular, we see that it is justified to ignore the term $\smash{\propto e^{(4+\alpha)ky}}$ of the dilaton vev in \eqref{prefactor}. Close to the UV brane, on the other hand, the prefactor is very small. This leads to a strong suppression of the kinetic term of the dark gauge boson in this region. As before, we therefore expect that the dark photon picks up only a small mass when the symmetry is broken at the UV brane. 

By expanding \eqref{dilatonaction} around $\langle \Phi \rangle$, we see that the action for fluctuations $\delta \Phi = \Phi - \langle \Phi \rangle$ of the dilaton is that of a massive bulk scalar with a mass on the UV brane. The KK modes of this scalar all have masses of the order $m_\ir$ and higher. We will therefore ignore them in the following.

We will now show that the spectrum indeed contains a light gauge boson. Using the decomposition 
\be
\label{gaugebosonKK}
A'_\mu (x^\nu,y) = \sum_{n=0}^{\infty} A'^{(n)}_\mu(x^\nu) f^{(n)}_\ai(y)\,,
\ee
the bulk equation of motion for the profiles follows from \eqref{dilatoncoupling}. In terms of the coordinate $z\equiv e^{ky}$, we have
\be
\label{lde}
\left[ \partial_z^2  + \left( \frac{\alpha \beta}{z^{\alpha+1}}- \frac{1}{z} \right) \partial_z + \frac{m_n^2}{k^2}\right] \, f_\ai^{(n)}(z) \, = \, 0 \, .
\ee
For simplicity, we focus on the case $\alpha=2$, corresponding to a dilaton with 5d mass $\smash{m_\Phi = \sqrt{12} k}$. To cast \eqref{lde} in the form of a Schr\"odinger equation, we make the ansatz $\smash{f_\ai^{(n)} = \sqrt{z} \, \exp(\frac{\beta}{2 z^2}) \tilde{f}_\ai^{(n)}}$. This gives
\be
\label{lde2}
\left[ \partial_{z}^2  +  \left( \frac{4 \beta}{z^4} -\frac{\beta^2}{z^6} - \frac{3}{4 z^2} + \frac{m_n^2}{k^2}\right) \right] \tilde{f}_\ai^{(n)} \, = \, 0 \, .
\ee
We see that, for light modes with masses $m_n \ll k$, the mass term can be neglected in the region close to the UV brane (where $\smash{z \sim 1}$). The resulting equation can be solved in closed form and we find
\be
\label{expint}
f_\ai^{(n)}(z) \, \simeq \, C_1 \, + \, C_2 \left(e^{\beta/z^2}z^2-\beta \Ei\left(\beta/z^2\right) \right) \, ,
\ee
where $\Ei$ is the exponential integral and $C_1, C_2$ are constants. As before, we break the gauge symmetry on the UV brane by imposing a Dirichlet boundary condition. Recall that we consider $\beta\gg1$, so using the fact that 
$\smash{\Ei(x) = e^{x}(x^{-1} + x^{-2} +\dots})$ for $x \gg 1$, we find
\be
\label{c1c2}
\frac{C_1}{C_2} \, \simeq \,   \frac{e^{\beta}}{\beta} \, .
\ee
This fixes the wavefunction in the UV (up to a normalization factor). In order to determine the mass spectrum, we need to continue this solution to the IR brane, where the wavefunction has to fulfill the boundary condition. Since $\Ei(x) \simeq \gamma + \log x $ for $x\ll 1$, the wavefunction \eqref{expint} for $\smash{z \gg \sqrt{\beta}}$ can be approximated by 
\be
\label{approxwf0}
f_\ai^{(n)}(z) \simeq  C_1 +  C_2 \, z^2 \, .
\ee
This agrees with the solution to \eqref{lde} for $\smash{z \gg \sqrt{\beta}}$ (in which case the term involving $\beta$ can be neglected) if one also neglects the mass term. We are looking for an exponentially light mode with mass $\smash{m_0 \ll k e^{-kL}}$. From \eqref{lde2}, we see that the mass term in this case is a small perturbation even at the IR brane where $z = e^{k L}$. Thus, we can solve the equation of motion as a perturbation series in $\smash{\frac{m_0^2}{k^2}}$. In particular, \eqref{approxwf0} is the solution (for large $z$) to zeroth order. Up to first order, we find
\be
\label{approxwf1}
f_\ai^{(0)}(z) \, \simeq \,  C_1 \left(1 \, + \, \frac{m_0^2}{k^2}\, \frac{z^2}{4} \left(1-2 \log z\right)\right)\, + \, C_2 \, \left(z^2 \, - \, \frac{m_0^2}{k^2} \frac{z^4}{8}\right) \, .
\ee
As before, we impose the Neumann boundary condition $\partial_z f_\ai^{(n)} |_{z_\ir} =0$ at the IR brane, corresponding to an unbroken gauge symmetry. 
Using \eqref{approxwf1} as well as \eqref{c1c2} and solving for $m_0$, we then see that the spectrum contains a mode with mass
\be
\label{smallmass}
m_0 \, \simeq \sqrt{\frac{2 \beta}{k L}} \, e^{-\beta/2} k \, .
\ee
For sufficiently large $\beta$, this mass is smaller than the IR scale (in which case the perturbative expansion, which led to \eqref{smallmass}, is justified).
In particular, for $m_\ir=1 \text{ TeV}$ and $k = 10^{18}$ GeV, a dark photon mass of $\mathcal{O}$(GeV) is obtained for $\beta \approx 84$. Recall that $\beta$ is given in terms of the boundary vev $v$ and the dilaton decay constant $f_\Phi$ by $\smash{\beta = 2 v/f_\Phi^{3/2}}$. The required $\beta$ can thus be obtained with the moderate hierarchy $v^{2/3} \approx 10 f_\Phi $ (where $ v^{2/3}$ has mass dimension 1). Alternatively, it is conceivable that the dilaton vev with the required coefficient $\beta$ can be generated without such a hierarchy in the parameters by using additional scalars with suitable bulk and boundary potentials.

We have determined the masses of the higher KK modes by solving \eqref{lde} numerically. The resulting spectrum is well approximated by the spectrum of massive KK modes of a gauge boson without bulk and boundary masses. In particular, the second lightest KK mode has a mass $m_1 \simeq 2.45 \, m_\ir$. Finally, we have also checked numerically that the mass of the lightest KK mode is well described by \eqref{smallmass}.

\subsection{Mixing with the Standard Model photon}\label{mixingsection}
In order to couple the dark photon to the electromagnetic current, we assume that the dark photon and the SM photon mix via a kinetic term. This mixing can be generated by integrating out some heavy states which are charged under both gauge groups. As we will discuss in more detail in Sect.~\ref{CouplingSection}, due to the dilaton factor, the coupling of the dark gauge boson to charged states grows towards the UV. It is therefore simplest to have these heavy states localized on the IR brane, leading to kinetic mixing on the IR brane. The relevant part of the action then reads:
\begin{equation}
S \, \supset \, \, \int d^5x \sqrt{-g} \left[ -\frac{1}{4}\, e^{-2\phi} F'_{MN} F^{\prime MN}  -  \frac{1}{4}\, F_{MN} F^{MN}  -  \frac{1}{ k} \,\zeta\, F'_{\mu \nu} F^{\mu\nu} \,\delta(y-L) \, \right]~,
\label{mixingaction}
\end{equation}
where $\zeta$ is a small dimensionless constant whose value depends on the details of the sector which generates the kinetic mixing and $\smash{F_{MN}^2}$ is the kinetic term of the SM photon (which we assume to be a bulk field). We have omitted a term $F^{\mu5}F'_{\mu 5}$ on the brane which vanishes as discussed in \cite{Carena:2002me}. Furthermore, for simplicity we do not consider boundary kinetic terms for the dark gauge boson and the SM photon. Such terms would not change our conclusions.

For simplicity, we restrict ourselves to the dilaton with a linear vev. As before, it is convenient to absorb the dilaton factor into the gauge field, $\widehat{A}_M = e^{-\langle\phi\rangle} A'_M$. Recall that we consider a dilaton vev which is small in the IR (cf.~Sect.~\ref{linDilaton}). The magnitude of the kinetic mixing term therefore only changes by an $\mathcal{O}(1)$ factor with this field redefinition. We absorb this factor into $\zeta$. Due to the kinetic mixing term, the KK decompositions of the dark gauge boson and the SM photon have to involve the same set of fields $A^{(n)}_\mu$. We thus decompose the fields as 
\be
\smash{\widehat{A}_\mu (x^\nu,y) = \sum_{n=0}^{\infty} f_\dark^{(n)}(y) A^{(n)}_\mu(x^\nu)} \quad \text{and} \quad  \smash{A_\mu (x^\nu,y) = \sum_{n=0}^{\infty} f_\sm^{(n)}(y)}A^{(n)}_\mu(x^\nu)~,
\ee
respectively. Imposing a Dirichlet condition on the UV brane for the dark gauge boson, its wavefunctions are again given by \eqref{massive}, with $\smash{f_\dark^{(n)} = \widehat{f}^{(n)}_\ai}$ and $N^{(n)}_\dark = N^{(n)}_\ai$. Similarly, imposing the Neumann condition on the UV brane for the SM photon, leads to
\be
 f_\sm^{(n)}(y)=N_\sm^{(n)} \, e^{ky}\left[
J_1\left(\frac{m_n}{k}e^{ky}\right)-\frac{J_0(\frac{m_n}{k})}{Y_0(\frac{m_n}{k})}
\; Y_1\left(\frac{m_n}{k}e^{ky}\right)\right]~.
\label{smphotonwf}
\ee
The presence of the kinetic mixing term modifies the boundary conditions at the IR brane which now mix the wavefunctions of the dark gauge boson and the SM photon:
\begin{gather}
\left(\partial_y  f^{(n)}_\dark -b k f^{(n)}_\dark - \zeta k^{-1} m_n^2 e^{2ky} f^{(n)}_\sm\right) \bigg|_{y=L}\, = \, 0~,\\
\left(\partial_y  f^{(n)}_\sm - \zeta k^{-1} m_n^2e^{2ky} f^{(n)}_\dark \right) \bigg|_{y=L}\, = \, 0~.
\end{gather}
This system decouples for a massless mode, $m_0=0$. Using also the boundary conditions at the UV brane, we see that the spectrum contains a massless KK mode $A^{(0)}_\mu$ with wavefunctions $f^{(0)}_\sm=\rm const.$ and $f^{(0)}_\dark=0$.
The quantization condition for the masses of the higher KK modes $\smash{A^{(n)}_\mu}$ reads
\be
\frac{J_{b}\bigl(\frac{m_n}{m_\ir}\bigr)+B_\dark
Y_{b}\bigl(\frac{m_n}{m_\ir}\bigr)}{J_{b-1}\bigl(\frac{m_n}{m_\ir}\bigr)+B_\dark
Y_{b-1}\bigl(\frac{m_n}{m_\ir}\bigr)} \, = \, -\zeta^2 \frac{m_n^2}{m_\ir^2} \frac{J_1\bigl(\frac{m_n}{m_\ir}\bigr)+B_\sm \,
Y_1\bigl(\frac{m_n}{m_\ir}\bigr)}{J_0\bigl(\frac{m_n}{m_\ir}\bigr)+B_\sm \,
Y_0\bigl(\frac{m_n}{m_\ir}\bigr)}~,
\label{massesboundarymixing}
\ee
where $\smash{B_\sm \equiv -J_0(\frac{m_n}{k})/Y_0(\frac{m_n}{k})}$ and $\smash{B_\dark \equiv -J_{b-1}(\frac{m_n}{k})/Y_{b-1}(\frac{m_n}{k})}$. By expanding the quantization condition for $\smash{m_n\ll m_\ir}$, assuming $\zeta <1 $ and $b>1$, we see that the spectrum again contains a light mode with mass given by \eqref{masseqn}. The masses of the higher KK modes can be determined numerically. For example when $\zeta=10^{-3}$ and $b = 1.2$, we find that the masses of the next two heavier KK modes are given by $m_2 \simeq 2.45 \, m_\ir $ and $m_3 \simeq 4.1 \, m_\ir $.

The boundary conditions also determine the ratio of the prefactors $N^{(n)}_\sm$ and $N^{(n)}_\dark$ in the wavefunctions $f^{(n)}_\sm$ and $f^{(n)}_\dark$:
\be
\frac{N^{(n)}_\sm}{N^{(n)}_\dark} \, = \, \zeta \frac{m_n}{m_\ir}\frac{J_{b-1}\bigl(\frac{m_n}{m_\ir}\bigr)+B_\dark
Y_{b-1}\bigl(\frac{m_n}{m_\ir}\bigr)}{J_0\bigl(\frac{m_n}{m_\ir}\bigr)+B_\sm \,Y_0\bigl(\frac{m_n}{m_\ir}\bigr)}~.
\label{normratio}
\ee
The overall normalization of the wavefunctions is determined by the following orthonormal condition, as can be seen from the action \eqref{mixingaction}:
\begin{multline}
\label{orthonormal}
\int_0^L dy \left[f^{(n)}_\sm(y) f^{(m)}_\sm(y)+ f^{(n)}_\dark(y)f^{(m)}_\dark(y) \right] \\ +\frac{\zeta}{k}\left( f^{(n)}_\sm(L) \, f^{(m)}_\dark(L) \, + f^{(m)}_\sm(L) \, f^{(n)}_\dark(L) \right)\,=\, \delta^{nm}~.
\end{multline}

The coupling of a KK mode $\smash{A^{(n)}_\mu}$ to KK modes which are charged under the dark gauge group is determined by an overlap integral involving the wavefunction $\smash{f^{(n)}_\dark}$. Similarly, the coupling to KK modes charged under the Standard Model $U(1)$ is determined by an overlap integral involving $\smash{f^{(n)}_\sm}$. The massless KK mode (with wavefunctions $\smash{f^{(0)}_\sm=\rm const.}$ and $\smash{f^{(0)}_\dark=0}$) therefore only couples to the electromagnetic current and not to the dark current. It can thus be identified with the 4d SM photon. The wavefunctions of the asymptotically light KK mode (with mass given by \eqref{masseqn}) can again be simplified by using the asymptotic forms of the Bessel functions. Using \eqref{normratio} and \eqref{orthonormal} to determine the normalization constants, we find that $\smash{f^{(1)}_\dark}$ is again well approximated by \eqref{lightwavefunction}, whereas $\smash{f^{(1)}_\sm}$ is approximately constant, $\smash{f^{(1)}_\sm \simeq \zeta /20}$ for $1<b <2$. We can compare the latter wavefunction with the normalized wavefunction of the 4d photon, $\smash{f^{(0)}_\sm \simeq 1/6}$. The coupling of the asymptotically light KK mode to the electromagnetic current is thus suppressed by the small mixing parameter $\zeta$ relative to the coupling of the 4d photon. This is analogous to the suppression in 4d when a gauge boson mixes with the SM photon via a small kinetic term (see \cite{Holdom:1985ag}). In order to estimate the coupling strengths of the heavier KK modes, we evaluate their wavefunctions at the IR brane, assuming $\zeta <1$. For the second massive KK mode, we find that $\smash{f^{(2)}_\sm(L)} \sim 1$ does not depend on $\zeta$, whereas $\smash{f^{(2)}_\dark(L)} \propto \zeta$. The third massive KK mode, on the other hand, has $\smash{f^{(3)}_\dark(L)} \sim 1$ and $\smash{f^{(3)}_\sm(L)} \propto \zeta$. When the mixing becomes smaller, the second mode therefore decouples from the dark current whereas the third mode decouples from the electromagnetic current. Accordingly, for vanishing mixing $\zeta=0$, these states are KK modes of the SM photon and the dark gauge boson, respectively.

\section{The Dark Matter} \label{secdark}
\subsection{Dark matter on the IR brane}
\label{IRDM}
As a simple realization of dark matter in our warped model, we can consider a fermion localized on the IR brane which is charged under U(1)$^\prime$ but neutral under the SM gauge group. 
The fermion thus couples to the dark photon with $\mathcal{O}$(GeV) mass and its higher KK modes. This realizes the model in \cite{dmref}. Due to the dark force, the annihilation of dark matter to dark photons in the galaxy halo is enhanced by the Sommerfeld effect. The resulting dark photons subsequently decay to the SM via the kinetic mixing discussed in Sect.~\ref{mixingsection}. Their mass is chosen such that decays to antiprotons are kinematically forbidden whereas decays to positrons (and electrons) are still allowed. This can give rise to the excess (over the expected background) in the combined flux of cosmic ray electrons and positrons which was observed by the experiments FERMI~\cite{fermi} and HESS~\cite{hess}.
It can also explain the rising fraction of cosmic ray positrons 
reported by the experiment PAMELA \cite{pamela} without producing antiprotons in excess of the observed flux. The best fits to the data are achieved with dark matter masses of the order TeV~\cite{Bergstrom:2009fa,mpsv} which is the natural scale for a fermion on the IR brane. 

Note that the annihilation of dark matter to the higher KK modes of the dark photon in the galaxy halo has to be forbidden or suppressed. As the latter have masses of $\mathcal{O}$(TeV) or higher, their decay would otherwise produce too many antiprotons. If the dark matter fermion is lighter than the lightest of these states (which has a mass $m_2 \simeq 2.45 \, m_\ir $ for the parameters used in Sect.~\ref{mixingsection}), such processes are kinematically forbidden.

Since the fermion couples to the dark photon which in turn couples to the electromagnetic current, strong constraints arise from direct detection experiments \cite{Pospelov:2007mp}. For a dark photon with mass of $\mathcal{O}$(GeV) and coupling $\alpha_\dark \sim 10^{-2}$ to the dark matter,
this requires the mixing $\zeta$ between the dark gauge boson and the SM photon (cf.~Sect.~\ref{mixingsection}) to be $\lesssim 10^{-6}$ \cite{Pospelov:2007mp,dmref}. The cross-section of dark matter scattering off nuclei mediated by higher KK modes of the dark photon is additionally suppressed by the TeV (or higher) mass in the propagator. These states therefore do not lead to a tighter constraint on $\zeta$. There are also lower bounds on $\zeta$. The requirement that the dark photon decays before nucleosynthesis leads to the bound $\zeta \gtrsim 10^{-11}$ \cite{Pospelov:2007mp}. Moreover, if the mixing is too small, the dark matter does not stay in thermal equilibrium with the SM until it freezes out. The corresponding bound on $\zeta$, however, requires a detailed analysis of the freeze-out process which is beyond the scope of this paper. 

Alternatively, the constraints from direct detection experiments can be evaded if the Dirac fermion is split into two Majorana states with different masses \cite{inelastic}. The dark photon couples dominantly off-diagonal to the two Majorana states. A scattering event in a direct detection experiment can therefore only occur if enough kinetic energy is available. In particular, by making the mass split larger, the event rate can be made sufficiently small. In addition, if the mass split is of order 100 keV, the annual modulation reported by the experiment DAMA/LIBRA \cite{dama} can be made consistent with the null results of other experiments~\cite{inelastic}.   

For a mass split of order 1 MeV, on the other hand, the heavier Majorana fermion decays to the lighter state via the emission of an electron-positron pair. Enhanced by the Sommerfeld effect, the heavier states are produced by scattering of the dark matter in our galaxy's halo. The resulting population of positrons (and electrons) with low injection energies can explain~\cite{exciting1a,exciting1b} the 511 keV line observed in the center of our galaxy, recently remeasured by the satellite INTEGRAL \cite{integral}. As the dark photon couples dominantly off-diagonal to the two Majorana states, however, the dominant excitation process is $\smash{\chi \chi \rightarrow \chi^* \chi^*}$ (where $\chi$ and $\smash{\chi^*}$ are the lighter and the heavier Majorana state, respectively). In particular, the process $\smash{\chi \chi \rightarrow \chi^* \chi}$ is strongly suppressed. For kinematical reasons, it can then be difficult to produce a sufficient amount of excited states (and thus positrons) \cite{dmref}. Nevertheless, it may be interesting to consider our model also for this purpose.

It would be interesting to see how a mass split can be generated for a fermion on the IR brane. Instead, in the next section, we will show how to induce a mass split for a fermion which lives in the bulk of the Randall-Sundrum model.

\subsection{A small mass split from localization}
\label{splitfromlocalization}

To split the fermion into two Majorana states, we consider a bulk Dirac fermion (instead of a fermion on the IR brane) and add a Majorana mass term on the UV brane. Small splittings between the Majorana components of the KK modes can be obtained by localizing the fermion away from the UV brane. 
The lowest lying KK state can be massive and therefore be automatically stable. This implements the proposal in \cite{dmref}, but lifted into the warped 5th dimension. 

Therefore, besides the dark photon, we introduce a dark sector bulk fermion $\chi=(\chi_\li, \chi_\ri)^T$ charged under U(1)$^\prime$ 
with a bulk Dirac mass term and a Majorana mass term on the UV brane. The action is given by
\begin{equation}
\label{fermionaction}
       S^{(\chi)} = - \int d^5x \sqrt{-g} \left[\frac{i}{2}\left(\bar\chi\Gamma^M D_M \chi -  \bar{D}_M\bar\chi\Gamma^M \chi\right)
       + i m_\chi \bar\chi\chi + i \delta(y) \frac{d'}{2} \left( {\bar \chi}^c\chi + \text{h.c.}\right)\right]~,
\end{equation}
where $\Gamma^M$ are the curved-space gamma matrices. Furthermore, the covariant derivative is given by ${D_M = \partial_M + i g_5^\prime A'_M + \omega_M}$, where $\smash{g_5^\prime = \mathcal{O}(k^{-1/2})}$ is the gauge coupling\footnote{On an orbifold, the gauge field $A'_M$ is odd under $\mathbb{Z}_2$ at the UV brane. In order for the gauge interaction term to be even under this $\mathbb{Z}_2$, the gauge coupling  then has to be proportional to the step function, $\smash{g_5^\prime} \propto \epsilon(y)$, around the UV brane. This is similar to the bulk mass $m_\chi$ being proportional to $\epsilon(y)$ on an orbifold.} and $\omega_M$ is the spin connection. We will discuss the coupling to the dark gauge boson in Sect.~\ref{CouplingSection}. We parameterise the bulk mass term as $m_\chi = c' k$ with a dimensionless constant $c'$
and the boundary Majorana mass term by the dimensionless constant $d'$. 
The Majorana term explicitly breaks the U(1)$^\prime$ symmetry which is consistent with the U(1)$^\prime$ symmetry being broken on the UV brane. Finally, note that we have not included a dilaton factor as in \eqref{dilatoncoupling}. If present, such a factor can be absorbed into $\chi$, leading again to \eqref{fermionaction}.

On the IR brane we impose $\chi_\ri|_{L}=0$ which is equivalent to $(\partial_y +(c'-2)k) \chi_\li |_{L}=0$.
Instead, the Majorana mass term on the UV brane leads to the boundary condition $\bar{\chi}_\ri|_0=\frac{d'}{2} \chi_\li|_0$ \cite{hs,Csaki:2003sh,tg}. We see that the spinor fields in the KK decomposition of $\chi_\li$ and $\chi_\ri$ can no longer be independent, corresponding to KK modes which are Majorana fermions. We thus perform the KK decomposition
\begin{equation}
\label{fermionKK}
\chi_\li(x^\mu,y) = \sum_{n} f_\li^{(n)}(y)\, \chi^{(n)}(x^\mu) \quad \text{and} \quad \chi_\ri(x^\mu,y) = 
\sum_{n}  f_\ri^{(n)}(y)\, \bar{\chi}^{(n)}(x^\mu)~,
\end{equation}
leading to the solutions
\begin{equation}
f_\pmi^{(n)}(y)=N^{(n)}\, S^{(n)}_\pmi \, e^{\frac{5}{2}ky} \left[ 
J_{\kappa_\pm} \left(\frac{|m_n|}{k}e^{ky}\right)-\frac{J_{\kappa_-}\left(\frac{|m_n|}{m_\ir}\right)}{Y_{\kappa_-}\left(\frac{|m_n|}{m_\ir}\right)}
\; Y_{\kappa_\pm}\left(\frac{|m_n|}{k}e^{ky}\right)\right]~,
\label{fermprofiles}
\end{equation}
where $\kappa_\pm \equiv c'\pm 1/2$ and $N^{(n)}$ is a normalisation constant (which is identical for the left- and right-handed wavefunction as can be seen using the equations of motion). Moreover, $\smash{S^{(n)}_\li \equiv \sgn(m_n)}$, where $\sgn$ is the sign function, and ${S^{(n)}_\ri\equiv1}$. As we will see, negative and positive mass eigenvalues are allowed and correspond to different physical states.

When $d'=0$, the massless zero-mode has the profile $f_\li^{(0)}\sim e^{(1/2-c') k y}$ with respect to a flat metric, and consequently can be localized anywhere in the bulk. This is analogous to the situation with bulk SM fermions~\cite{gn,gp1}. However, when $d' \sim1$ and $\smash{c' > -\frac{1}{2}}$, this mode has a non-negligible coupling to the UV brane, and will feel the large 
boundary Majorana mass. Consequently it obtains a mass much larger than $m_\ir$, thereby decoupling from the low-energy 
spectrum. In this case the lightest states become the lowest-lying KK modes which have Dirac masses of order $m_\ir$. Since the KK modes 
have only a small wavefunction overlap with the UV brane, they will only receive a small Majorana mass and via the seesaw mechanism, the former Dirac states obtain a small mass split. Therefore, we expect to find a KK tower of two nearly degenerate (pseudo-Dirac) mass eigenstates.

The mass spectrum is easily obtained by imposing the UV brane boundary condition on the profiles (\ref{fermprofiles}). This leads to the condition
\begin{equation}
\frac{J_{\kappa_-}\bigl(\frac{\smash{|m_n|}}{m_\ir}\bigr)}{Y_{\kappa_-}\bigl(\frac{\smash{|m_n|}}{m_\ir}\bigr)} \, = \, \frac{J_{\kappa_-}\left(\frac{\smash{|m_n|}}{k}\right) \mp \frac{d'}{2} \, J_{\kappa_+}\left(\frac{\smash{|m_n|}}{k}\right) }{Y_{\kappa_-}\left(\frac{\smash{|m_n|}}{k}\right) 
\mp \frac{d'}{2} \, Y_{\kappa_+}\left(\frac{\smash{|m_n|}}{k}\right)}~,
\label{massquantization1}
\end{equation}
where the signs correspond to $\smash{\sgn(m_n)=\pm}$, respectively.
By expanding the Bessel functions with small arguments $\smash{\frac{|m_n|}{k} \ll1}$, we find that the right-hand side of \eqref{massquantization1} for $d' \sim 1$ is of order $\smash{\bigl(\frac{|m_n|}{k}\bigr)^{2c'}}\hspace{-1mm}$, which is a small number for $c'>0$. The mass eigenvalues to lowest order $\smash{m_n^{(0)}}$ are therefore determined by 
\begin{equation}
   J_{\kappa_-}\left(\frac{\smash{|m_n^{(0)}|}}{ m_\ir }\right) \, = \, 0~.
\end{equation}
This is solved by $\smash{|m_n^{(0)}| \simeq (n + \frac{c'-1}{2}) \pi \, m_\ir}$ for $n=1,2,\dots$. Note that for every positive solution
$\smash{m_n^{(0)}}$ there is also a negative solution, $\smash{-m_n^{(0)}}$, which together represent a Dirac fermion. 
We will denote the corresponding mass eigenvalues with negative indices, such that $\smash{m_{-n}^{(0)}\equiv -m_n^{(0)}}$. Similarly, $\smash{\chi^{(-n)}}$ and $\smash{f_\pmi^{(-n)}}$ denote the corresponding KK states and their wavefunctions, respectively.

At the next order these two mass eigenstates are split by the boundary Majorana mass. By expanding \eqref{massquantization1} around $\smash{m_n^{(0)}}$, the leading correction is found to be 
\begin{equation}
 m_n^{(1)} \, \simeq \, -\frac{2\pi}{d'} \frac{1}{\Gamma(\frac{1}{2}+c')^2} \left(\frac{|m^{(0)}_n|}{2 k}\right)^{2c'}\, m_\ir~. 
\label{splitmass}
\end{equation}
Since for $d' \sim 1$ and $c'>0$, the leading correction is $\smash{|m_n^{(1)}| \ll |m_n^{(0)}|}$, a small mass split is induced in the mass
eigenstates $\smash{|m_n| \simeq |m_n^{(0)} + m_n^{(1)}|}$. We therefore indeed find a KK tower of two nearly degenerate Majorana states.
Assuming $k = 10^{18}$ GeV and $m_\ir = 1$ TeV, the mass splitting is $\mathcal{O}(\text{MeV})$ for $c'\simeq 0.22$ and thus in the right range to explain the INTEGRAL signal (see however the caveat discussed in Sect.~\ref{IRDM}). For $c'\simeq 0.26$, on the other hand, the mass splitting is $\mathcal{O}(100 \text{ keV})$ and in the range relevant to DAMA/LIBRA. 
The actual dark matter is identified with the lowest lying KK state $\chi^{(1)}$. For $c'$ in the range $0.22- 0.26$,
the corresponding dark matter mass is $m_\dm \simeq (1.96-2.02) \, m_\ir$. Thus, in our model, the dark matter is naturally in the TeV range. 

Finally, recall that (see Sect.~\ref{IRDM}) the annihilation of dark matter into higher KK modes of the dark gauge boson has to be forbidden or suppressed. For the parameters used in Sect.~\ref{mixingsection}, the lightest of these KK modes has a mass $m_2 \simeq 2.45 \, m_\ir $. Since $m_\dm < m_2$ for the mass of dark matter in the above range, these processes are kine- matically forbidden.

\subsection{Coupling to the dark photon}
\label{CouplingSection}
We will now discuss the coupling of the dark matter to the dark gauge boson in some detail. Using the KK decomposition for the fermion \eqref{fermionKK} and for the dark gauge boson \eqref{gaugebosonKK}, the kinetic term in \eqref{fermionaction} gives
\be
\label{interaction}
S\, \supset \,  -i \int d^4x \, \left(\,  \sum_{r,s}  \,Z_{rs}\, \bar{\chi}^{(r)} \, \bar{\sigma}^\mu \partial_\mu \, \chi^{(s)} \, + i \sum_{r,s,\ell} \, g_{rs\ell}\, \bar{\chi}^{(r)} \, \bar{\sigma}^\mu A'^{(\ell)}_\mu \, \chi^{(s)}  \right) \, .
\ee
We have to require that $\smash{Z_{rs} \equiv \int dy \, e^{-3ky} ( f^{(r)}_\li f^{(s)}_\li + f^{(r)}_\ri f^{(s)}_\ri)} = \delta_{rs}$ to obtain canonical kinetic terms. Due to the boundary Majorana mass, the fermion wavefunctions $f^{(n)}_\li$ and $f^{(n)}_\ri$ can only be orthonormalized in a generalized sense (see e.g.~\cite{Casagrande:2008hr}). Using the equations of motion, we find that we have to impose
\be
\label{norm}
\begin{split}
\int dy \, e^{-3ky} f^{(r)}_\li f^{(s)}_\li \,& = \, \frac{1}{2} \delta_{rs} + \,  \frac{d'}{2} \, \frac{f^{(r)}_\li(0)f^{(s)}_\li(0)}{m_r+m_s}\, , \\
\int dy \, e^{-3ky} f^{(r)}_\ri f^{(s)}_\ri \,& = \, \frac{1}{2} \delta_{rs} - \,   \frac{d'}{2} \, \frac{f^{(r)}_\li(0)f^{(s)}_\li(0)}{m_r+m_s} \, ,
\end{split}
\ee
which gives $Z_{rs}=\delta_{rs}$. The coupling constants $g_{rs\ell}$ are given by the overlap integral of the fermion and gauge boson wavefunctions:
\be
g_{rs\ell} \, = \, g_5^\prime \, \int dy \, e^{-3 ky} \left(\,f_\li^{(r)} \, f_\li^{(s)}  f_\ai^{(\ell)} - f_\ri^{(r)} \, f_\ri^{(s)}  f_\ai^{(\ell)} \,\right) \,.
\label{couplingprime}
\ee

We will first evaluate the couplings $g_{rs0}$ involving the dark photon. As we have discussed in Sect.~\ref{lightgeneral}, for Neumann boundary conditions at the two branes, the action \eqref{ZeroModeAction} allows for a massless mode with constant profile, $\smash{f_\ai^{(0)}(y) = N_\ai^{(0)}}$. 
If we instead impose the Dirichlet condition at the UV brane (and choose a suitable dilaton profile), the resulting light mode still has an approximately constant profile: 
For the exponential dilaton, this follows from the fact that $\smash{f_\ai^{(0)}(y) \propto  e^\beta/\beta + e^{2ky}}$  according to \eqref{c1c2} and \eqref{approxwf0}. From \eqref{smallmass}, we see that $\smash{e^\beta/\beta \gg e^{2kL}}$ for a light mode with ${m_0 \ll m_\ir}$. The nonconstant part of the wavefunction, $\smash{\propto e^{2ky}}$, is then negligible even at the IR brane. For the linear dilaton, we have redefined the field variable and the wavefunctions, ${\widehat f}_\ai^{(n)} \equiv e^{-\langle \phi \rangle} f_\ai^{(n)}$. According to \eqref{lightwavefunction} the wavefunction of the light mode then is ${\widehat f}_\ai^{(0)} \propto e^{bky}$. Going back to the original field variable, we see that the wavefunction is again (approximately) constant, $\smash{f_\ai^{(0)}(y) \simeq N_\ai^{(0)}}$. Therefore using this approximation and substituting \eqref{norm} into the overlap integral \eqref{interaction}, we find
\be
\label{zmcoupling}
g_{rs0} \; \simeq \; g_4^\prime \,d' \; \frac{f^{(r)}_\li(0)f^{(s)}_\li(0)}{m_r+m_s} \; \sim \; \begin{cases}
                                                                                \quad \; \; g_4^\prime    &\;\,\text{$r=-s$} \\
                                                                                \frac{g_4^\prime}{\max(|r|,|s|)} \left(\frac{|m_r| \,|m_s|}{k^2}\right)^{c'}  
                                                                                &\text{otherwise,}
                                                                               \end{cases}
\ee
where we have defined the dimensionless constant $g_4^\prime \equiv N_\ai^{(0)} g_5^\prime$. Using the normalized wavefunction \eqref{lightwavefunction} and the dilaton profile as discussed at the end of Sect.~\ref{linDilaton}, we find $\smash{N_\ai^{(0)} \simeq \sqrt{2 b k}}$. Similarly, for the exponential dilaton, requiring a canonical kinetic term for the light mode from \eqref{ZeroModeAction}, we find $\smash{N_\ai^{(0)} \approx 1/\sqrt{L}}$. Thus, for $\smash{g'_5 \sim k^{-1/2}}$, we have $g'_4 \sim 1$.
In the last step in \eqref{zmcoupling}, we have expanded the wavefunction \eqref{fermprofiles} for $\smash{\frac{|m_n|}{k} \ll 1}$ and used \eqref{massquantization1} and \eqref{splitmass}. We see that the largest coupling is between two Majorana states with nearly degenerate mass (i.e. pairs $\smash{r=-s}$ whose masses are split by \eqref{splitmass}). All other couplings are heavily suppressed. In particular, interactions coupling a Majorana state to itself (i.e.~for $r=s$), are suppressed by a factor of the order $\smash{m^{(1)}_r/m_r}$ (where $\smash{m^{(1)}_r}$ is the mass split \eqref{splitmass}).
The same suppression factor arises in 4d when a Dirac fermion is split into two Majorana states via a small Majorana mass (see e.g.~\cite{inelastic}). The fact that the coupling is dominantly off-diagonal between two quasi-degenerate Majorana states implements the inelastic dark matter scenario \cite{inelastic} to reconcile DAMA/LIBRA with other direct detection experiments.

\subsection{Coupling to higher modes of the dark gauge boson}

We will now discuss couplings involving the higher KK modes of the dark gauge boson, focusing first on the linear dilaton. 
For simplicity, we will neglect the mixing with the SM photon and use the unmixed wavefunctions for the dark gauge boson. 
Using the dilaton profile as discussed at the end of Sect.~\ref{linDilaton}, the coupling constants in terms of the redefined wavefunctions $\smash{{\widehat f}_\ai^{(n)}}$ read
\be
\label{couplingtilde}
g_{rs\ell} \, = \, g_5^\prime \, \int dy \, e^{-3 ky}e^{bk(L-y)}\left(\,f_\li^{(r)} \, f_\li^{(s)}  {\widehat f}_\ai^{(\ell)} - f_\ri^{(r)} \, f_\ri^{(s)}  {\widehat f}_\ai^{(\ell)} \,\right)~.
\ee
Recall that, in the rescaled field variable $\widehat{A}_M$, the action of the dark gauge boson becomes that of a `standard' gauge boson in a Randall-Sundrum model (but with bulk and boundary masses). We now see that the coupling $\smash{\hat{g}_5(y) \equiv g_5^\prime e^{bk(L-y)}}$ of this gauge field is $y$-dependent. Moreover, for $b > 1$ (which leads to an exponentially light dark photon), this gauge coupling grows towards the UV brane.
This is also visible in the field variable $A'_M$: In order to localize the dark photon towards the IR, the prefactor of the kinetic term in \eqref{dilatoncoupling} has to decay towards the UV. This prefactor is related to the inverse gauge coupling and we therefore again find that the gauge coupling grows in the UV. As heavier KK modes live closer to the UV brane, one can be worried that the KK modes will eventually become strongly coupled. We will now show that this is not the case.

To this end, we approximate the wavefunctions using the expansions of Bessel functions for small and large arguments. For example, for $f^{(n)}_\li$ this gives
\be 
f^{(n)}_\li(z) \, \sim \,   \sqrt{\frac{m_\ir m_n}{k}} z^{5/2} \begin{cases} 
                                                                        \left(\frac{m_n}{k}z\right)^{-1/2} \hspace*{-.1cm}  \left( \cos\left(\frac{m_n}{k}z\right) +\left(\frac{m_n}{k}\right)^{2c^\prime} \hspace*{-.1cm} \cos\left(\frac{m_n}{k}z\right)\right) & \text{$\frac{m_n}{k}z \gg1$}, \\
                                                                        \; \quad \quad \left(\frac{m_n}{k}z\right)^{c'+1/2} +\left(\frac{m_n}{k}\right)^{2c^\prime} \left(\frac{m_n}{k}z\right)^{-c'-1/2}  & \text{$\frac{m_n}{k}z \ll1 ,$} 

                                                                        \end{cases}
\ee
where $\smash{z\equiv e^{ky}}$ and we have neglected phases in the trigonometric functions. From this expansion, we see that $\smash{|f^{(n)}_\li(y)|}$ is everywhere smaller than or of the order $\smash{\sqrt{m_\ir} \, e^{2ky}}$. Similarly, 
we find that 
\be
|f^{(n)}_\ri(y)| \, \lesssim \, \sqrt{m_\ir} \, e^{2ky} \qquad \text{and} \qquad |{\widehat f}^{(n)}_\ai(y)| \, \lesssim \, \sqrt{m_\ir} \, e^{ky/2} \,. 
\ee
Using these estimates in \eqref{couplingtilde}, we obtain the following upper bound on the gauge couplings:
\be
\label{gce}
|g_{rs\ell}| \; \lesssim  \; g_5^\prime \, m_\ir^{3/2}  e^{bkL} \int dy \, e^{(3/2-b)ky} \, \sim \, g_4^\prime~.
\ee
The last step is valid for $\smash{b<\frac{3}{2}}$ (recall that we chose $b=1.2$), where we have used the fact that $\smash{N_\ai^{(0)} \sim \sqrt{k}}$. We thus find that, for $\smash{g_4^\prime \lesssim 1}$, all gauge couplings involving higher KK modes of the dark gauge boson are perturbative. This is due to the fact that the overlap integral is dominated by the IR. Note that \eqref{gce} is only an upper bound. In particular, we can expect that the gauge couplings involving heavy KK modes are much smaller than 1 due to the fact that oscillations of the wavefunctions (which we have neglected in \eqref{gce}) cancel out each other.

Let us now consider the gauge couplings for the exponential dilaton, focusing on the case $\alpha=2$. As in Sect.~\ref{expDilaton}, we will use the frame in which the dilaton factor is not absorbed into the wavefunction. In this frame, there is no $y$-dependent coupling in the overlap integral which has the form (\ref{couplingprime}).
We will first consider modes with masses $\smash{m_n \ll k / \sqrt{\beta}}$. From the Schr\"odinger equation \eqref{lde2}, we see that terms depending on $\beta$ (coming from the dilaton background) can be neglected for $\smash{z \gg \sqrt{\beta}}$. One can check that the dark gauge boson fulfills the same equation of motion as the SM photon in this region. Correspondingly, the wavefunctions $\smash{f_\ai^{(\ell)}}$ have the same profile as KK modes of the SM photon (see \cite{dhr,pomarol}). Imposing the Neumann condition at the IR brane, we find
\be
f_\ai^{(\ell)}(z) \; \simeq \; N_\ai^{(\ell)} \, e^{ky}\left[
J_1\left(\frac{m_n}{k}e^{ky}\right)-\frac{J_0(\frac{m_n}{m_\ir})}{Y_0(\frac{m_n}{m_\ir})}
\;Y_1\left(\frac{m_n}{k}e^{ky}\right)\right] \qquad \text{for }z \gg \sqrt{\beta}~.
\label{BesselSol}
\ee
In the region $\smash{z \ll k / m_n}$, on the other hand, the mass term can be neglected. If we impose the Neumann condition (instead of a Dirichlet condition) at the UV brane, we therefore find from \eqref{lde}:
\be
\label{constsol}
f_\ai^{(\ell)}(z) \, \simeq \, \text{const.}  \qquad \text{for }z \, \ll \, k/m_n~.
\ee
By expanding \eqref{smphotonwf}, we see that the KK modes of the SM photon have approximately constant profiles in the region $\smash{z \ll k/m_n}$ as well. For modes with masses $\smash{m_n \ll k / \sqrt{\beta}}$, the regions overlap where \eqref{BesselSol} and \eqref{constsol} are valid. Thus, for the Neumann condition at the UV brane, the wavefunctions $\smash{f_\ai^{(\ell)}}$ have similar profiles as KK modes of the SM photon everywhere between the two branes. Accordingly, KK modes of the dark gauge boson couple with similar strengths to charged states as KK modes of the SM photon. By imposing a Dirichlet condition at the UV brane instead, the wavefunctions $\smash{f_\ai^{(\ell)}}$ are additionally suppressed in the UV, resulting in slightly smaller coupling strengths.

We will now discuss modes with masses $\smash{m_n \gtrsim k / \sqrt{\beta}}$. For sufficiently heavy modes, the mass term completely dominates the potential of the effective Schr\"odinger equation \eqref{lde2}. Imposing the Dirichlet condition at the UV brane, we then find
\be
\label{wfheavy}
f_\ai^{(\ell)}(z) \, \simeq \, N_\ai^{(\ell)} e^{\beta/2z^2} \sqrt{z} \, \sin\Bigl(\frac{m_n}{k} (z-1) \Bigr)~.
\ee
We see that the wavefunction is enhanced by the factor $\smash{\exp (\langle\phi\rangle) = \exp(\beta/2z^2)}$. This factor, which enters the overlap integral \eqref{couplingprime} via the wavefunction \eqref{wfheavy}, corresponds to the factor $\smash{\exp(\langle\phi\rangle)}$ coming from the $y$-dependent coupling in \eqref{couplingtilde}. By requiring a canonical kinetic term when inserting the profile in \eqref{dilatoncoupling}, we find $\smash{N_\ai^{(\ell)}\sim \sqrt{m_\ir}}$. Using the approximate wavefunction \eqref{wfheavy} and the upper estimates for $\smash{f^{(n)}_\li}$ and $\smash{f^{(n)}_\ri}$, we can evaluate the overlap integral. We then find that, for $\beta \lesssim 110$ assuming $\smash{kL \approx 34}$, the couplings are again perturbative.

Note that the perturbativity of the 4d couplings does not yet prove that loop corrections are small. For a given loop diagram, one has to sum over all KK modes of the intermediate fields. As a gauge theory in 5d is nonrenormalizable, a cutoff $\Lambda$ on the loop integrals has to be imposed. For example, one can introduce a Pauli-Villars regulator field with mass $\Lambda$. As discussed in \cite{Pomarol:2000hp}, the KK modes of the Pauli-Villars field pair up with the KK modes of the fields in the loop. Each pair leads to a finite contribution to the sum which can then be evaluated. We will not perform this analysis here. But as we have seen, for the exponential dilaton, modes of the dark gauge boson with masses up to the order $k/\sqrt{\beta}$ have similar wavefunctions as those of the SM photon. This corresponds to the fact that the dilaton background for the exponential vev becomes relevant only close to the UV brane. As loop corrections due to the SM photon can be expected to be small, we believe that corrections due to these states are small as well. Larger corrections could then only arise from heavy states with masses between $k/\sqrt{\beta}$ and $\Lambda$. We furthermore emphasize that the dark sector is gauge invariant up to the breaking at the UV brane.

Let us briefly discuss loop effects from the 5d viewpoint. In the frame in which we absorb the dilaton factor into the gauge field, the gauge coupling becomes large (in units of $\smash{k^{-1/2}}$) in the UV.  Note, however, that the propagators of the dark gauge boson and the dark matter fermion are suppressed in the UV due to their bulk and boundary masses. This gives a compensating effect to the growing coupling in the UV. More precisely, note that the exponential dilaton profile can be localized more towards the UV brane by choosing larger $\alpha$ (cf.~Sect.~\ref{expDilaton}). Although we have shown the existence of a light mode only for the case $\alpha=2$, we expect that such a mode appears in the spectrum for larger $\alpha$ as well. Absorbing the dilaton factor into the gauge field, the $y$-dependent gauge coupling for the exponential dilaton becomes
\be
\widehat{g}'_5(y) \; = \; e^{e^{- \alpha k y} \beta/2} \, g'_5~,
\label{expdilcoup}
\ee
where $\smash{g_5' \sim k^{-1/2}}$  is the gauge coupling appearing in the covariant derivative in \eqref{fermionaction}. In the formal limit $\alpha \rightarrow \infty$, the $y$-dependent factor in \eqref{expdilcoup} is one everywhere in the bulk and jumps to a large value only at the UV brane. Due to the Dirichlet boundary condition, the gauge field however vanishes at the UV brane. Thus, at least in the formal limit $\alpha \rightarrow \infty$, there appears to 
be no issue with the UV coupling.\footnote{Alternatively note that in string theory, examples are known where the dilaton coupling diverges in the UV such as the near-horizon geometry of D4-branes~\cite{imsy}. In these cases there is a dual UV description and we may also speculate that a similar dual description exists for our field theory model as well.}

In any case, loop corrections for the scenario with dark matter in the bulk deserve a further analysis.

\subsection{Freeze-out}

At temperatures above the IR scale, the IR brane in a Randall-Sundrum model is replaced by a black hole horizon~\cite{ahpr}. This corresponds to 
the dual gauge theory being in the deconfined phase for temperatures above its confinement scale. As the Universe cools a phase transition to the 
Randall-Sundrum phase takes place (corresponding to the confinement phase transition of the dual gauge theory) 
at or somewhat below the IR scale~\cite{ncr}. 

Since the freeze-out temperature of a thermal relic with mass $m_\dm$ is typically of the order $m_\dm/20$, we assume that the details of the phase transition can be neglected and perform a standard freeze-out calculation for our dark matter (see also \cite{as}). Recall from Sect.~\ref{CouplingSection} that the dark photon couples dominantly to nearly degenerate partners $\smash{\chi^{(n)}}$ and $\smash{\chi^{(-n)}}$. All other couplings are strongly suppressed. Such a suppression does not occur for couplings involving the higher KK modes of the dark gauge boson. If produced during the phase transition, higher KK modes $\chi^{(n)}$ of the fermion decay via these states to $\chi^{(1)}$ or $\chi^{(-1)}$ plus SM particles. For a mode with mass $m_n$, the corresponding decay rate is
\be
\Gamma \, \sim \, \alpha \, \alpha_\dark \zeta^2 \, m_n \, ,
\ee
where $\alpha$ is the fine-structure constant, $\alpha_\dark=g_4^{\prime 2}/ 4 \pi$ and $\zeta$ parametrizes the mixing with the SM photon. For $\alpha_\dark \sim 10^{-2}$ and $\zeta \sim 10^{-3}$, these decays happen on short timescales compared to the Hubble time during the phase transition and thus before the dark matter freezes out. 
The decay of $\smash{\chi^{(-1)}}$ to $\smash{\chi^{(1)}}$ plus SM particles, however, happens after freeze-out due to the kinematic suppression resulting from the small mass split.
The dominant annihilation channel for the dark matter is $\chi^{(1)}\chi^{(1)} \rightarrow \gamma^\prime\gamma^\prime$, where the dark photons subsequently decay to the SM. In this process (see Fig.~\ref{annihilationFIG}), all the KK modes $\chi^{(n)}$ can be exchanged. However, as only $\chi^{(-1)}$ couples unsuppressed to $\chi^{(1)}$ and the dark photon, the exchange of all other KK modes can be neglected.\footnote{For the same reason, we can also neglect coannihilations of $\chi^{(1)}$ with $\chi^{(-1)}$.} The annihilation cross section during freeze-out is thus given by
\begin{equation}
    \langle \sigma v\rangle_{\rm freeze} \, \sim \, \frac{\alpha^2_\dark}{m_\dm^2}~.
    \label{integral}
\end{equation}
As the dark matter particles still have high velocities during freeze-out, there is no significant Sommerfeld enhancement of the annihilation.\footnote{As was recently shown in \cite{Zavala:2009mi,Dent:2009bv}, however, in certain extreme cases the annihilation can be enhanced by an order of magnitude. In these cases, a correspondingly lower tree-level cross section has to be chosen in order to obtain the right relic abundance.} 
The enhancement turns on later during structure formation \cite{kp} in galaxy halos, when the dark matter is sufficiently slow. The right relic abundance can be obtained for ${m_\dm = \mathcal{O}(\text{TeV})}$ and $\smash{\alpha_\dark \sim 10^{-2}}$.

\section{Conclusion} \label{secDiscuss}
We have presented a warped model of the dark matter sector that incorporates a nonsupersymmetric
solution to the gauge hierarchy problem. The dark matter sector contains a WIMP near the TeV scale
that annihilates predominantly into leptons via a dark force. 
This realizes the scenario in Ref.~\cite{dmref} which was recently put forth to explain
cosmic ray anomalies observed by the experiments PAMELA, FERMI and HESS. 
In this scenario, the annihilation of dark matter is a two-step process: The dark matter 
first annihilates into dark photons which subsequently 
decay to the Standard Model. In order to allow for this decay while increasing the
annihilation cross section of dark matter via the Sommerfeld effect, 
the dark gauge group must be broken at a low scale, chosen to be of order GeV. 
In our warped model, the dark gauge group
is broken at the UV brane. The mass hierarchy between the Planck and GeV scale is then obtained by 
localizing the dark photon near the IR brane via a coupling to a dilaton background. Moreover,
in order to allow for the decay to the Standard Model, the dark photon is assumed to mix with the Standard Model photon.
In our model, this is realized by a kinetic mixing term which is localized on the IR brane.

We have furthermore showed how to introduce a small mass splitting for the dark matter in our model,
by including a Majorana mass on the UV boundary (where the dark gauge group is broken). This leads
to a pair of nearly degenerate (pseudo-Dirac) states with a tiny mass splitting.
The dark photon couples dominantly off-diagonal to these two states, avoiding constraints from
direct detection experiments if the mass splitting is sufficiently large. Moreover, for a mass splitting of order MeV, 
the results of DAMA/LIBRA can be reconciled with the null results of other experiments via the inelastic dark matter scenario.

Our model crucially depends on a dilaton background which enters via a prefactor into the 
kinetic term of the dark gauge boson. We provided a dynamical solution 
that showed how a suitable profile for the dilaton arises from boundary potentials. 
The prefactor of the gauge kinetic term is of order one in the IR and decays towards the UV, leading to the localization of the 
dark photon near the IR brane. But this also means that the effective 5d coupling of the dark gauge boson,
being related to the inverse of this factor, grows 
in the UV. 
However, the exponentially suppressed profiles of the dark matter and the dark gauge boson in the 
UV do lead to sufficiently weak coupling between the 4d Kaluza-Klein states. 
Nevertheless, loop corrections deserve a further analysis. 

Alternatively, the dark matter (and all other matter charged under the dark gauge group) can be 
localized at the IR brane, where the 5d coupling is not large. It would be interesting to see how a small mass 
splitting for the dark matter can be induced in this setup. Alternatively, constraints from direct detection experiments 
can be fulfilled by making the mixing between Standard Model photon and dark photon sufficiently small.

Finally, the dual 4d holographic description requires the entire 
dark sector to emerge at the TeV scale as the bound states of some unknown strong dynamics. The 5d warped model 
therefore provides a suitable low energy description with which to study a strongly-coupled dark sector
in a framework that easily incorporates and addresses the gauge hierarchy and fermion mass hierarchy
problems in the Standard Model.

\section*{Acknowledgments}
We thank Brian Batell, Emilian Dudas, and Nicholas Setzer for helpful discussions. This work is supported by the Australian Research Council.

\end{document}